\documentclass[a4paper]{article}
\pdfoutput=1

\usepackage[english]{babel}
\usepackage{algpseudocode}
\usepackage{multirow}
\usepackage[normalem]{ulem}
\usepackage{chemformula}
\usepackage{upgreek}
\usepackage{chemmacros}
\usepackage[autostyle]{csquotes}
\usepackage{url}
\usepackage{comment}
\usepackage{authblk}
\usepackage{natbib}
\usepackage{booktabs}
\usepackage{siunitx}
\usepackage{csquotes}
\usepackage{mathtools}
\usepackage{tabularx}
\usepackage[colorlinks = true,
            linkcolor = blue,
            urlcolor  = blue,
            citecolor = blue,
            anchorcolor = blue]{hyperref}

\makeatletter
   \def\@citecolor{blue}
   \def\@urlcolor{blue}%
   \def\@linkcolor{blue}%

\def\orcidID#1{\href{http://orcid.org/#1}{\protect\includegraphics{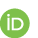}}}
\makeatother

\newcommand{\ECnum}[4]{EC #1.#2.#3.#4}
\newcommand{\uniprot}[1]{UniProtKB: #1}
\newcommand{\mcsaid}[1]{M-CSA ID: #1}
\newcommand{\rhea}[1]{RHEA:#1}

\newcommand{\dextro}{{\fontencoding{T1}\fontshape{sc}\selectfont d}}
\newcolumntype{R}{>{\raggedleft\arraybackslash}X}

\date{}

\begin{document}
\title{Graph Transformation for Enzymatic Mechanisms}

\author[1]{Jakob L. Andersen \orcidID{0000-0002-4165-3732}}
\author[1]{Rolf Fagerberg \orcidID{0000-0003-1004-3314}}
\author[2]{Christoph Flamm \orcidID{0000-0001-5500-2415}}
\author[3]{Walter Fontana \orcidID{0000-0003-4062-9957}}
\author[1]{Juraj Kol\v{c}\'ak \orcidID{0000-0002-9407-9682}}
\author[1]{Christophe V.F.P. Laurent \orcidID{0000-0002-9112-6981}}
\author[1]{Daniel Merkle \orcidID{0000-0001-7792-375X}}
\author[1]{Nikolai N{\o}jgaard \orcidID{0000-0002-7053-4716}}

\affil[1]{Department of Mathematics and Computer Science, University of Southern Denmark, Odense, Denmark}
\affil[2]{Department of Theoretical Chemistry, University of Vienna, Vienna, Austria}
\affil[3]{Department of Systems Biology, Harvard Medical School, Boston, Massachusetts, USA}

\maketitle

\begin{abstract}
    \textbf{Motivation:}
	The design of enzymes is as challenging as it is consequential for making chemical synthesis in medical and industrial applications more efficient, cost-effective and environmentally friendly. While several aspects of this complex problem are computationally assisted, the drafting of catalytic mechanisms, i.e.\@ the specification of the chemical steps---and hence intermediate states---that the enzyme is meant to implement, is largely left to human expertise. The ability to capture specific chemistries of multi-step catalysis in a fashion that enables its computational construction and design is therefore highly desirable and would equally impact the elucidation of existing enzymatic reactions whose mechanisms are unknown.\\
    \textbf{Results:}
	We use the mathematical framework of graph transformation to express the distinction between rules and reactions in chemistry. We derive about 1000 rules for amino acid side chain chemistry from the M-CSA database, a curated repository of enzymatic mechanisms. Using graph transformation we are able to propose hundreds of hypothetical catalytic mechanisms for a large number of unrelated reactions in the Rhea database. We analyze these mechanisms to find that they combine in chemically sound fashion individual steps from a variety of known multi-step mechanisms, showing that plausible novel mechanisms for catalysis can be constructed computationally.\\
    \textbf{Availability and Implementation:}
	The prototype implementation sources are available at
	\href{https://github.com/Nojgaard/mechsearch}{https://github.com/Nojgaard/mechsearch}. \\
    A live demo is accessible at \href{https://cheminf.imada.sdu.dk/mechsearch/}%
    {https://cheminf.imada.sdu.dk/mechsearch/}. \\
    \textbf{Contact:}
    \href{mailto:daniel@imada.sdu.dk}{daniel@imada.sdu.dk}
    \\
    \textbf{Supplementary information:} Supplementary data are available at \\
    \href{https://cheminf.imada.sdu.dk/preprints/ECCB-2021}%
    {https://cheminf.imada.sdu.dk/preprints/ECCB-2021}
\end{abstract}


\section{Introduction}
\label{sec:introduction}

Since the advent of the digital revolution, the vast repertoire of chemical knowledge has become accessible through a growing number of repositories. The utility of such warehousing hinges on computational tools for searching and aggregating content and for exploring its consequences. Indeed, many tools exist for reasoning about molecules and their reactions at the level of symbolic chemistry \citep{Todd:2005,Cook:2012,Segler:2017}. Likewise, software implementing quantum and classical methods enables the study of configurational energy landscapes that undergird symbolic chemistry \citep{Welborn:2019}.

The toolbox dwindles, however, when it comes to reasoning about the chemistry of reaction \emph{networks}. While computational and mathematical infrastructure exists for studying the kinetics or flux balance of networks, there is little in the way of systematically \emph{constructing} such networks while taking into account chemical possibilities.
Tools aimed at the kinetics of chemical reaction networks are cast in terms of variables that refer to concentrations, but their dynamics alone cannot introduce new components beyond those explicitly specified at the outset. To extend a network requires an executable representation of actual chemistry.

The construction, and therefore also the design, of chemical networks is made possible by the notion of a chemical rule, which is distinct from a chemical reaction. In a reaction, molecules are completely specified, whereas a rule makes explicit only those aspects of molecules that are necessary for a reaction to occur at the level of abstraction defined by symbolic chemistry. A rule is a schema that represents the transformation of an educt pattern into a product pattern. Given completely specified educt molecules, a rule generates possible reactions for those molecules that contain its educt pattern. Since symbolic chemistry represents molecules as typed graphs, the formal domain of graph transformation \citep{EhrigPS1973,HabelMP2001,EhrigEGT2006} seems to be the natural foundation for implementing the distinction between reactions and rules.

The distillation of rules from large catalogs of reactions would open the door to the iterative construction and design of networks by repeated application of specifically chosen rules. Rule collections with a formally sound application semantics make chemical knowledge \enquote{executable}, but the realization of this notion in the form of computational tools is challenging.

In this contribution we provide an initial example towards realizing this vision. A key component is an open-source software platform, known as M{\O}D, for specifying and iteratively applying chemical rules \citep{AndersenFMS2013,mod}. We deploy this platform on rules that we derive from a database of hand-curated mechanisms of enzymatically catalyzed reactions known as the \enquote{Mechanism and Catalytic Site Atlas} or M-CSA \citep{RibeiroHFTFT2017}. 

Our specific focus, thus, is on the design of enzymatic catalysis. Such design is of significance in a range of applications from addressing disease to shifting chemical industry towards more sustainable, waste minimizing, and environmentally friendly production processes \citep{Zimmerman:2020,Pleissner:2020,Schrittwieser:2018}. Chemical networks are central in this goal because almost all catalysis rests on a network-based mechanism, despite informal language often referring to a singular agent (the catalyst). Specifically, at the catalytic site of an enzyme several reaction steps combine into a network in such a way that upon completion of the overall transformation from substrates into products each protein component has regained the same chemical state it had initially. This requires the network to be a cycle. Cyclical network catalysis also occurs at larger scales. For example, the citric acid cycle at the core of modern biological metabolism acts as a network catalyst regardless of the fact that its individual steps are also catalyzed by enzymes.

Designing a full enzyme requires attention to structure, which controls specificity and provides a stable niche that guarantees a proper causal ordering of the catalytic steps. While the network view does not address structure, it underscores that designing an enzyme also requires designing a multi-step catalytic \emph{process}. The computational implementation of this view through graph transformation should lend further credence to the computer assisted design of enzymes \citep{Welborn:2019}.

Our contribution is organized as follows. We first proceed by making a subset of the M-CSA executable as graph transformation rules. We then exemplify the utilization of such rules by constructing proposals for catalytic network mechanisms for some reactions in the Rhea database \citep{Lombardot2019}---a database unrelated to the M-CSA, also containing enzymatic reactions, but mostly listed without suggestions for an underlying network mechanism. 

\section{Methods and Materials}
\label{sec:materials_methods}
\subsection{Enzyme Catalysis and the M-CSA}
\label{sec:catalysis}

Like any other reaction, an enzymatic reaction is usually expressed as the conversion of educt molecules (substrates) into products by means of a specific protein functioning as catalyst. Biology would quite literally be unthinkable without catalysis. For example, the spontaneous decarboxylation of arginine occurs roughly with a rate constant of \SI{2e-17}{\per\second} (a half-life of about 1.1 billion years), whereas the \textit{Escherichia coli} arginine decarboxylase has a reported \num{7e19} fold catalytic proficiency \citep{Snider2000}.

Enzymatic catalysis, however, is oftentimes not a single event but involves multiple steps that together constitute a catalytic mechanism. Each of these steps can be seen as an elementary reaction in which components of the substrate, amino acids, and possibly cofactors (such as flavin adenine dinucleotide), react to a stable intermediate state that becomes the input to the subsequent step, eventually resulting in the formation of the product and regeneration of the catalyst. While the packaging of a cyclic reaction network within a large protein warrants referring to enzymatic catalysis as if it were a single event, it is essential for our purpose to unpack catalysis into a detailed mechanism supporting the overall reaction. This mechanism not only transforms a substrate into a product but must also guarantee that any protein components deployed in the process regain their initial state upon completion.

For clarity we fix terminology as follows. The phrase \enquote{reaction step} (or \enquote{step} for short) is used to denote a reaction judged \emph{elementary}, that is, further indivisible, by M-CSA contributors. Moreover, we use the phrase \enquote{reaction mechanism} to refer to a causal succession of steps and the phrase \enquote{overall reaction} as the chemical sum over a mechanism. \enquote{Reaction} can mean any of the above, depending on the context.

The \enquote{reaction center} of a step refers to the atoms that undergo an electronic displacement or whose bonds are rearranged by the reaction step (boldface atoms in \hyperref[fig:rule_derivation]{Figure \ref{fig:rule_derivation}}); see also \hyperref[sec:graph_transformation]{Section \ref{sec:graph_transformation}}. We refer to everything else that stays fixed and is not in the reaction center as the \enquote{context} of a step.

We further classify an amino acid explicitly mentioned in a step as either active or passive depending on whether it intersects the reaction center or not. Although passive amino acids are not subject to bond rearrangements, they are nonetheless deemed critical for efficient catalysis by contributing to the physicochemical properties of the catalytic pocket, such as establishing charge distributions and spatial constraints.

The M-CSA \citep{RibeiroHFTFT2017} contains overall reactions for a collection of $964$ representative enzymes (so-called \emph{entries} in M-CSA terminology). $684$ of these reactions are listed with at least one manually curated step-by-step description of the catalytic mechanism that converts reactants into products. The validity of the mechanisms considered for inclusion is judged on the basis of direct experimental evidence and observations explained by it. This can result in the inclusion of several reaction mechanisms for the same enzymatic overall reaction. In total, the M-CSA provides $818$ detailed reaction mechanisms.

The reaction mechanisms in the M-CSA include an English description, based on direct or indirect experimental evidence, of the catalytic process in terms of active and passive amino acids as well as any cofactors involved. A reaction step is formally expressed as an arrow pushing diagram describing the displacement of electrons. \hyperref[fig:rule_derivation]{Figure \ref{fig:rule_derivation}A} depicts a step as obtained from the M-CSA.

The M-CSA aims at providing a representative (non-redundant) set of all known enzymatic reactions, hence resulting in a collection of mechanisms that are considerably dissimilar.
Yet, many of the individual steps across different mechanisms appear similar or outright identical to each other if one were to restrict the context. This suggests that enzymatic mechanisms are composed of identifiable building blocks best described by rules at the step level. Once identified, such rules could be used to construct mechanisms for reactions other than those in the M-CSA.

\subsection{Graph Transformation Framework}
\label{sec:graph_transformation}

\begin{figure*}
\centering
\includegraphics[width=\textwidth]{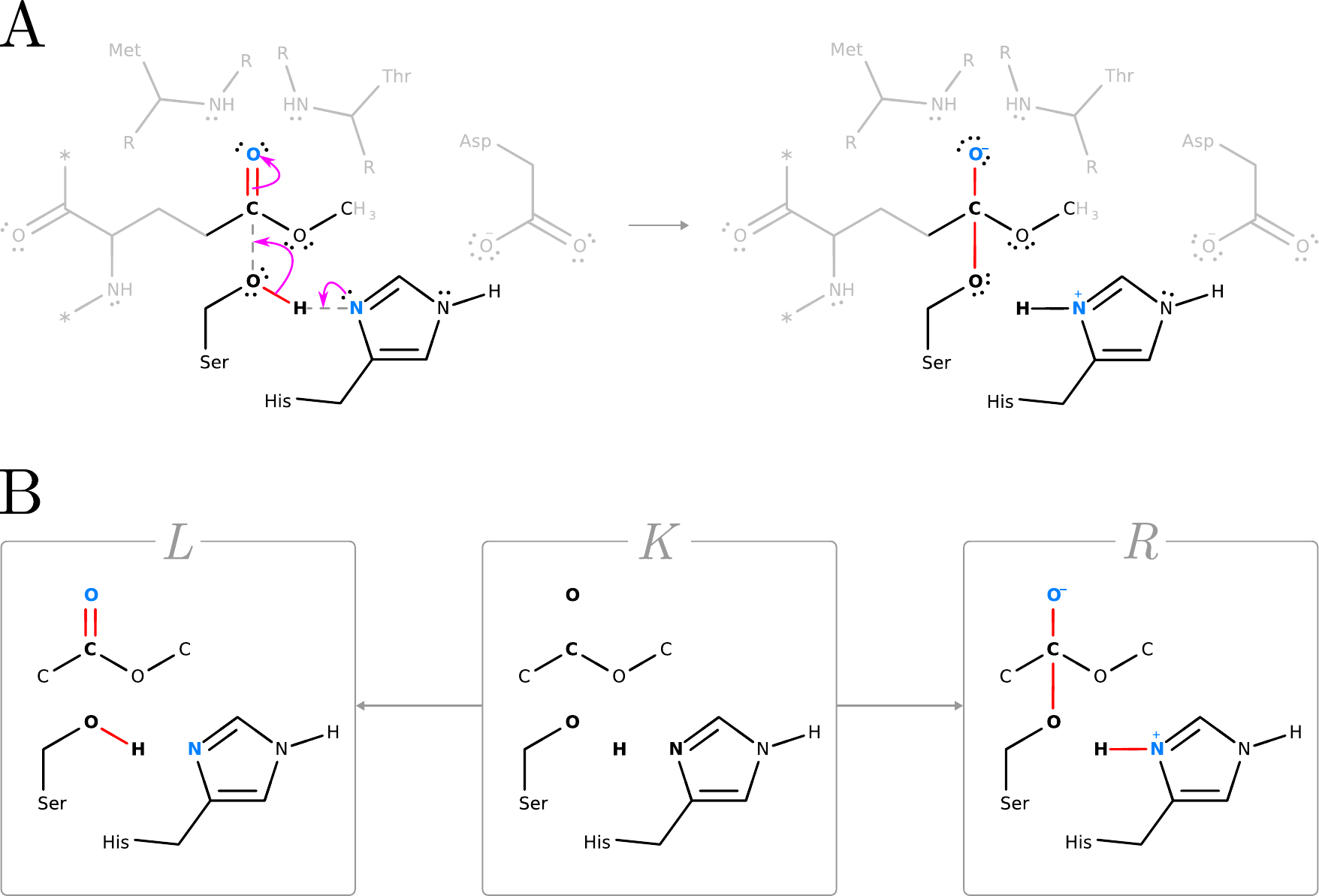}
\caption{%
	The relationship between a chemical step and a graph transformation rule. %
	(\textbf{A}) %
	The panel is adapted from step 1 in mechanism 1 of \mcsaid{337} and shows the initial reaction step performed by the protein-glutamate methylesterase (CheB) (\uniprot{P04042}, \ECnum{3}{1}{1}{61}, \ECnum{3}{5}{1}{44}). Aspartate increases the \pKa~of the histidine imidazole ring by forming a hydrogen bond to the histidine \ch{N_{$\epsilon$}}. This turns histidine into a powerful base and hence activates the serine residue. The latter is then able to perform a nucleophilic attack on the glutamine methyl-ester substrate. The left-hand side illustrates the electron movement from the histidine \ch{N_{$\delta$}} atom via serine to the \ch{O} atom in the carbonyl in the substrate ester group. The right-hand side of the arrow shows the resulting intermediates.	Electronic displacements are shown as arrow pushes in magenta. For details on gray shades refer to the caption of panel B. 
	(\textbf{B}) %
	This panel shows the graph transformation rule derived from panel A following our heuristic guidelines (\hyperref[sec:rule_contruction]{Section \ref{sec:rule_contruction}}). Explicit hydrogen atoms have been added for clarity. The bonds and nodes that change from $L$ to $R$, are emphasized in color (red for bonds, blue for nodes). The reaction center is shown in boldface. The rule asserts that those parts that are grayed out in panel A constitute molecular elements that have no bearing on the chemical transformation.	Structures were downloaded from the M-CSA.}
\label{fig:rule_derivation}
\end{figure*}

To make reaction knowledge executable, a formal framework is needed. Since for many purposes symbolic chemistry represents molecules as connected graphs that are undirected and typed, the formalism of graph transformation (or graph rewriting) appears to be an appropriate choice. Graph transformation is an extension of the idea of term rewriting to graphs and has a well-developed foundation in category theory \citep{EhrigPS1973, EhrigEGT2006}.

In a molecule graph, a typed node represents an atom with a charge and a typed edge represents a bond of a certain order (single, double, triple, aromatic). A collection of molecules as a whole is then represented as a disconnected graph, whose connected components represent the individual molecules. We refer to such a graph as a \enquote{state graph} (or \enquote{state} for short). Graph transformation is about defining rules by which one state can be transformed into another. The idea of a rule is to specify the transformation of a graphical input pattern into an output pattern and to carry this transformation over to the state if it contains the input pattern. 

In more formal terms, a rule (a double-pushout rule in the jargon of graph rewriting) is a span $p = (L\xleftarrow{l}K \xrightarrow{r}R)$, where $L$ and $R$ indicate the left and right pattern, respectively. $K$ is the invariant graph, containing elements common to $L$ and $R$. The correspondence between atoms in $L$ and $R$ (the atom map) is specified by the injections $l$ and $r$. The transformation from $L$ into $R$ is then given by breaking those bonds in $L$ that are not in $K$ and forming those bonds in $R$ that are not in $K$. The atoms in $L$ and $R$ also carry modifiable state, such as charge (blue font in \hyperref[fig:rule_derivation]{Figure~\ref{fig:rule_derivation}B}). 

A rule must at the very least specify all the bonds that are broken and formed in $L$ and $R$, respectively, or that undergo order modification. We call such a minimal rule the \emph{action}. The action is the formalization of the reaction center mentioned in the previous section. A \enquote{maximal rule} (or \enquote{reaction rule}) is one in which $L$ and $R$ are completely specified educt and product molecules. The \emph{context} $C$ is given by all nodes and bonds that refine (add invariant detail to) the action. An action, thus, has an empty context and a maximal rule contains as much context as needed for $L$ and $R$ to specify molecular species. By varying the context between these two extremes we can construct a variety of rules that are all refinements of the same action.

Recall that a state $G$ is a typically disconnected graph whose components are molecules. A rule transforms one state into another by \emph{application}. A rule application consists in embedding $L$ into the host graph $G$ and replacing the subgraph of $G$ selected by the embedding with $R$, while respecting the atom map given by $l$ and $r$. For example, in \hyperref[fig:rule_derivation]{Figure \ref{fig:rule_derivation}}, the rule shown in panel B is applied to the molecules on the left of the reaction arrow in panel A, yielding the depicted reaction. By applying a rule to a state in general and not just to state components it actually modifies (the educts) we let the state act as a context for successive applications of rules, i.e.\@ a mechanism. This will simplify how we present our strategy in \hyperref[sec:methodology]{Section \ref{sec:methodology}}.

\subsection{Converting M-CSA to Graph Transformation Rules}
\label{sec:rule_contruction}

Like any reaction, a step can be viewed as the special case of a maximally refined rule. Such a rule consists of all molecular species, including passive ones, that the M-CSA curators deemed to be necessary for the complete documentation of the step. Maximal rules are probably overspecified because the context includes molecular parts that are unlikely to all be necessary for setting off the chemical transformation. In particular, rules that mirror reactions are unlikely to overlap with one another in a way that permits their composition. Their $L$-patterns are unlikely to be embeddable in anything other than the molecular species represented by $L$ itself.

By decreasing the context $C$ of a maximally refined rule we might better capture the chemistry that drives a reaction and decrease the chemical specificity of a rule in a fashion that provides more opportunities for composition. This would facilitate the construction of reaction networks. Clearly, by going to the extreme of emptying $C$, thus retaining the action alone, we might misrepresent the chemistry and increase compositionality too indiscriminately. 

Given a collection of partially redundant reaction examples, a major challenge is to devise statistical methods for identifying the right amount of context to be enshrined in a rule that abstracts a reaction class. The M-CSA, however, is not the right collection for such an endeavor because its objective is to provide maximal non-redundant coverage of distinct enzymatic mechanisms.

For the present purpose we use, therefore, heuristics that are also aimed at keeping the combinatorial explosion of the state space generated by the repeated application of rules in check. Since educt combinations of molecular species often contain upwards of $100$ atoms, we require a relatively large context to keep the state space (\hyperref[sec:methodology]{Section \ref{sec:methodology}}) manageable. At the same time, we try to limit the extent of context---especially context originating from the substrate of an enzymatic reaction---by invoking patterns commonly used in chemistry. These considerations led us to devise three guidelines for crafting the context $C$ of a step. We then assemble a rule by adding $C$ to the action.

\begin{enumerate}
\item \textbf{Local topology:}
    We assume that the immediate surroundings of the reaction center are a significant driver of the reaction. Hence, all atoms and bonds that are directly connected to the reaction center are retained in the context. For instance, the immediate surroundings allow us to distinguish between reactions acting on a carbon chain or a methyl group.
\item \textbf{Functional patterns:}
    We compiled a set of $157$ chemical patterns based on functional groups, cycles, and small molecules common in organic chemistry (e.g.\@ carboxylic acid, imidazole, water). In addition, the set is adapted to the M-CSA by including minor variations in the patterns that were observed in the substrates utilized by the enzymes documented in the M-CSA.
    The chemical patterns are then embedded into both the reactant and product graphs of a reaction step. If a match intersects the reaction center, all the atoms and bonds of the pattern are included in the invariant graph $K$.
    A list of chemical patterns can be found in the
    \href{https://cheminf.imada.sdu.dk/preprints/ECCB-2021}%
    {supplementary data}.
\item \textbf{Active amino acids:}
    We posit that the active amino acids are crucial for driving a reaction step. If any part of an amino acid intersects the reaction center, the whole amino acid side-chain is included in the context.
\end{enumerate}

A reaction step in arrow pushing notation and the rule inferred on the basis of these guidelines is shown in \hyperref[fig:rule_derivation]{Figure~\ref{fig:rule_derivation}}.

Our rules do not include any molecules that do not share at least one atom with the reaction center. This may well misrepresent the chemistry, because aspects of these molecules could be necessary for allowing the reaction to proceed. For example, they might act as electrostatic stabilizers, shift \pKa~values, provide hydrogen bonds or guide the reaction sterically. None of these properties can be represented explicitly in the graph transformation framework. They could, however, be represented implicitly, precisely by including the molecular parts responsible for these properties in the $L$ and $R$ patterns. Through the $L$ pattern they act as necessary (matching) conditions for the rule to apply. Although the level of abstraction defined by graph transformation cannot represent the physical causes of a reaction directly, it can be informed by them. This suggests that molecular dynamics, quantum mechanical calculations, or more phenomenological approaches, including machine learning, that are capable of determining the molecular patterns required for a reaction will be useful in defining the content of a rule. By shaping a rule, this information becomes executable. However, an augmentation of rule construction of this sort is beyond the present scope. Here, we simply argue heuristically.

The proposed coarsening of a reaction step into a rule can be readily applied to any step in the M-CSA. Because the objective of the M-CSA is to non-redundantly cover known enzymatic reaction mechanisms, its mechanisms are diverse and include steps that are fairly complex and perhaps overly specific or rare. This does not serve well the creation of a rule set that could be applied to find catalytic mechanisms for a variety of reactions. Moreover, our present framework cannot handle some of the complexity present in the M-CSA and we therefore separate out those we can handle.

First, we exclude reaction steps that rely on metal ions as a cofactor. Metal ions frequently interact with molecules through coordination bonds. This type of covalent bond comprises two electrons originating from only one atom. At the current state of development, our graph model for chemistry is unable to represent the chemistry of coordination bonds.

Second, we exclude reaction steps from mechanisms that rely on radicals and single electron jumps. This kind of electron movement is also frequently associated with metal ion chemistry.

The presentation of mechanisms in the M-CSA targets human readers. As a consequence, across the presentation of steps of a mechanism some molecules might appear, disappear, or change abstraction level. While such editing is beneficial for the purpose of visualization, it is detrimental for rule construction. In particular, it hampers correct tracking of individual amino acids in the event of covalent bond formation with the substrate. Thus, thirdly, we submit all mechanisms to a sanity check by tracking atoms across the full complement of molecules mentioned across their steps. Many of the changes are easy to detect and fix by propagating the disappearing or appearing molecules across the mechanism. However, if our atom tracking fails, we exclude all the steps of the mechanism.

Taking this filtering into account, we obtain a total of $1083$ different rules derived from reaction steps across $471$ M-CSA mechanisms. For $368$ of those mechanisms, all steps are used for rule construction, making them fully reproducible by our rule set. The majority of disqualified reaction steps is dependent on metal ions. We consider this limitation acceptable, especially since we focus on the chemistry of catalysts that do not require cofactors.

We can link the rule set thus obtained to the chemical process classification tags of each step provided by the M-CSA. Each of the five most common chemical processes in the M-CSA is instantiated by at least \SI{49}{\percent} of reaction steps used in the construction of our rule set. Specifically, proton transfer is represented with \SI{58.8}{\percent}, proton relay with \SI{49.2}{\percent}, unimolecular elimination by conjugate bases with \SI{68.5}{\percent}, bimolecular nucleophilic substitutions and additions are both represented with \SI{62.8}{\percent}.

\subsection{Using Rules to Propose Novel Mechanisms}
\label{sec:methodology}

In the previous section we described our procedure for converting a reaction step in the M-CSA into a graph transformation rule. In this section we focus on how a collection of rules so obtained can be used to reconstruct M-CSA mechanisms or propose new ones for reactions not included in the M-CSA.

Recalling \hyperref[sec:graph_transformation]{Section \ref{sec:graph_transformation}}, the term \enquote{state} refers to a graph
where each connected component represents a molecule. Any particular molecule can occur in multiple instances to accommodate stoichiometry. Moreover, since reactions conserve mass, all states have an invariant atomic composition. We first need to define the states that are reachable with a given set of rules $\mathcal{R}$ from an initial state $G_0$. For this we need a bit of notation.

For a rule $p\in\mathcal{R}$ to apply to a state $G$, the left-hand pattern $L$ of the rule must embed in $G$. We refer to such an embedding as a \enquote{match} and define the set of all possible matches of a given rule in state $G$ as $M(p, G)$. We write $G\xRightarrow{p, m}H$ for the transformation of $G$ into $H$ by rule $p\in\mathcal{R}$ using the match $m\in M(p,G)$. Each such rule application is a \enquote{direct derivation} which we call a \emph{transition} between states. Recall from \hyperref[sec:graph_transformation]{Section \ref{sec:graph_transformation}} that $G$ is the full complement of molecules as it might have been formed by prior rule applications. It may include molecules that are not altered by rule $p$. A state $G$ can thus be thought of as the test-tube mixture in the context of which a rule application (a reaction step) takes place to produce a new state $H$. Restricting a state to only components that are matched by $m$ then corresponds to the usual notion of a reaction (a \enquote{proper derivation} \citep{strat:14}).

Aggregating all $H$ that can be generated by all rules in $\mathcal{R}$ using all possible matches onto $G$, we obtain the set of states that constitute a $1$-step extension of $G$. In symbols, we define the $k$-step extension of $G_0$ as
\begin{align*}
\mathcal{G}_0 &= \left\{G_0\right\} \\
\mathcal{G}_{k+1} &= \left\{H\,\middle|\, \exists G\in\mathcal{G}_k, p\in \mathcal{R}, m\in M(p, G) : G\xRightarrow{p,m}H\right\}\cup\mathcal{G}_0
\end{align*}
$\mathcal{G}_k$, together with the transitions for all $p$ and $m$, define the reachable \emph{state space} $S_k$ at depth $k$. Specifically, $S_k$ is a directed graph with the node set $\mathcal{G}_k$ and an edge $(G, H)$ between nodes $G, H\in \mathcal{G}_k$ if the transition $G\xRightarrow{p, m}H$ exists for some rule $p$ and match $m$.

A further notion is that of a trace $\tau$, which is a sequence of transitions, that transforms the initial educt state $G_I\equiv G_0$ into the final product state $G_F$ within a state space $S$. A trace thus carves a path from $G_I$ to $G_F$ in $S$; it corresponds to a (match-)specific composition of rules that generate the transitions in $\tau$. We write $G_I\xRightarrow{\tau}G_F$ to denote the transformation of state $G_I$ into state $G_F$ by $\tau$. Since transitions represent reaction steps, $\tau$ represents a mechanism for transforming the educt molecules in $G_I$ into product molecules in $G_F$. In principle, a trace could take many detours in going from $G_I$ to $G_F$. In the present setting it makes sense to eliminate such meandering by stipulating that $\tau$ be a minimal path from $G_I$ to $G_F$ in $S$.

Since we aim at catalytic reactions we require certain molecules, represented by graph $A$, to be a subgraph of \emph{both} $G_I$ and $G_F$. To line up with intuition, we write $G_I=E\oplus A$ and $G_F=P\oplus A$, where $\oplus$ is the disjoint graph union. We then speak of an $A$-catalyzed transformation trace $\tau_A$ of educts $E$ into products $P$, in symbols $E \xRightarrow{\tau_A}P$. Our task, therefore, is to identify a possible mechanism $\tau_A$ within the state space $S_k$ constructed from the initial state $G_I$. The restriction to $S_k$ means that the mechanisms cannot be longer than $k$ steps.

The difficulty, of course, is in searching for potential mechanisms in a state space $S_k$ whose size is beyond astronomic even for modest $k$. Although the M-CSA contains mechanisms of up to $20$ steps in length, these appear to be outliers. Most M-CSA mechanisms are relatively short, averaging $3.4$ steps with a median of $3$. It seems justifiable, therefore, to limit the state space $S_k$ to some small iteration depth $k$. 

In order to improve our chances of finding a $\tau_A$ we leverage the information in both $E$ and $P$ when constructing the state space $S_k$. To find a $\tau_A$ such that $E\xRightarrow{\tau_A}P$ at depth $k$, the construction of $S_k$ starts from $G_I$ and we make this explicit by writing $S_k(G_I)$. However, we need to be certain that the state $G_F$ is contained in the node set of $S_k(G_I)$, that is, $G_F$ must be reachable from $G_I$. We can ensure this by exploiting the invertibility of graph transformation rules in the double pushout framework \citep{10.1007/BFb0025714}. Specifically, we invert all rules $p\in\mathcal{R}$ by swapping their left-hand $L_p$ and right-hand pattern $R_p$, to obtain an action that transforms the pattern $R_p$ into $L_p$. We then join $S_k(G_F)$, after inversion of all transitions, with $S_k(G_I)$ on the shared parts of the underlying states to obtain a state space that is guaranteed to contain all paths from $G_I$ to $G_F$ of length at most $2k$, while only exploring each state space to a depth of $k$.

Such a combination still results in a graph with possibly numerous \enquote{dead ends}, that is, states that are reachable only from $G_I$ or only from $G_F$. We remove such states in order to obtain a succinct representation and refer to the resulting state space as the \emph{relevant} state space. The relevant state space can be envisioned as a flow with a single source $G_I$ and a single drain $G_F$.


\section{Results}
\label{sec:results}
The approach described in \hyperref[sec:materials_methods]{Section~\ref{sec:materials_methods}} has been implemented in Python with the more computationally intensive tasks in C++. In constructing a state space we rely on MØD \citep{mod} for efficient graph transformations and on NetworkX \citep{networkx} for general graph algorithms. During the conversion of steps to rules, we use the Marvin Molecule File Converter 20.20.0 (ChemAxon Ltd., \url{https://chemaxon.com/}) for adding explicit hydrogen atoms to the Marvin files downloaded from the M-CSA database. MarvinSketch 20.20.0 (ChemAxon Ltd.) was used to draw the molecules in the figures.

To test the practicality of our approach, as well as the generality of the constructed rule set, we use reaction data provided by Rhea \citep{Lombardot2019}. Rhea is an expert-curated database containing information about reactions of biological interest, many of which are enzymatic and obtained from peer-reviewed literature.

Most Rhea reactions are not annotated with detailed catalyst information. While one or more proteins might be mentioned, the catalytic sites are generally unknown or not reported. The reactions provided by Rhea are therefore suitable for testing our ability to propose new mechanisms using the rules we constructed from the M-CSA.

The choice of catalysts for a reaction is part of the mechanism prediction process. In general, enzymatic reactions rely on a combination of amino acid side chains and possibly cofactors. As mentioned in \hyperref[sec:rule_contruction]{Section \ref{sec:rule_contruction}}, to curtail the combinatorial explosion we limit ourselves to catalysis that employs only amino acids. To this end, we identify $26$ tautomers of $17$ amino acids commonly used within the M-CSA database (\href{https://cheminf.imada.sdu.dk/preprints/ECCB-2021}{supplementary data}). Alanine, glycine and proline were not included.

\subsection{Single Amino Acid Mechanisms for Rhea Reactions}
\label{sec:result_stats}

\begin{table}
    \centering
    \begin{tabularx}{\linewidth}{lRRRRRR}
    	\toprule
    	               & Glu & His & Asp & Cys & Lys & Thr \\ \midrule
    	Rhea           & 624 & 421 & 226 &  50 &   5 &   1 \\
    	Rhea (unique)  & 135 &  28 &  73 &   4 &   4 &   1 \\
    	M-CSA (global) & 579 & 660 & 621 & 280 & 407 & 157 \\
    	M-CSA (single) &  25 &  34 &  20 &   7 &  14 &   4 \\
    	\bottomrule
    \end{tabularx}
    \vskip6pt
    \caption{Comparison of amino acids utilization frequency. The first row specifies the total number of reactions for which a mechanism using the specified amino acids was discovered. The second row counts the reactions for which all the discovered mechanisms relied on the particular amino acid. The third row lists total number of incidences across the entire M-CSA database. The fourth row counts all the M-CSA mechanisms represented by our rule set that only use a single active amino acid, of the type specified.}
    \label{tab:used_amino_acid}
\end{table}

\begin{figure*}
    \centering
    \includegraphics[width=\textwidth]{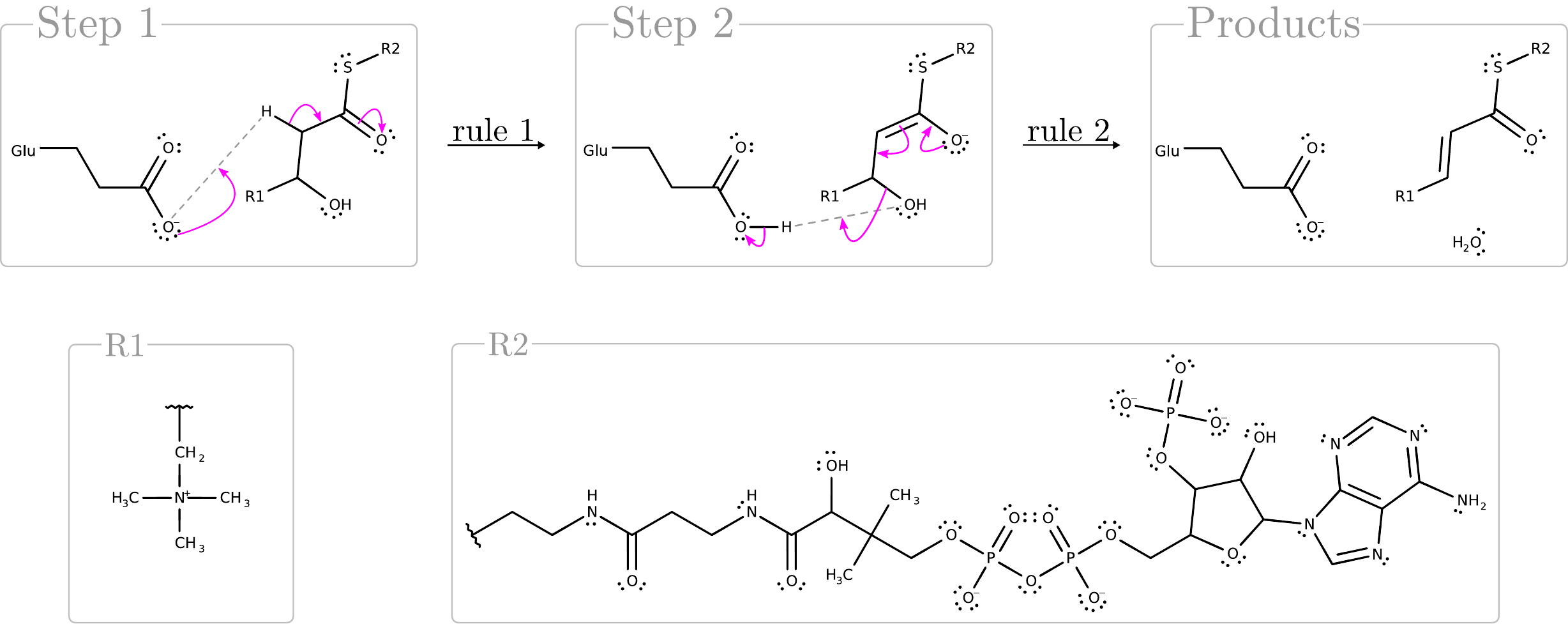}
    \caption{%
    Proposed 2-step reaction mechanism for Rhea reaction \rhea{28338}. %
    Application of rule 1 (derived from \mcsaid{341}, mechanism proposal 1, step 1 and \mcsaid{499}, proposal 1, step 1) causes a hydrogen abstraction from the substrate by a glutamate residue which causes the formation of an oxyanion. %
    The application of rule 2 (derived from \mcsaid{341}, mechanism proposal 1, step 1 and \mcsaid{947}, proposal 1, step 2) results in the collapse of the oxyanion, deprotonation of glutamate and the release of a water molecule. %
    Electronic displacements are shown as arrow pushes in magenta.%
    }
    \label{fig:Crotonobetainyl-CoA_hydratase}
\end{figure*}

The Rhea database lists over \num{e4} reactions. Some of these cannot be analyzed at the level of abstraction of our present model, such as conversions of substrates between tautomers differing only sterically. After removing these, we are left with $8805$ reactions for which we attempt to predict mechanisms within the limits of computational complexity.

We refer to the $368$ mechanisms in the M-CSA that are covered by our rule set as the \enquote{covered} M-CSA mechanisms. \SI{35.3}{\percent} of the covered M-CSA mechanisms utilize only a single active amino acid, suggesting that a comprehensive analysis of potential mechanisms employing just one amino acid tautomer across the $8805$ reactions from Rhea could be of interest.

The covered M-CSA mechanisms exhibit an average number of steps of $3.47$ with a median of $3$, similar to the set of all M-CSA mechanisms. Moreover, only $29$ (\SI{7.9}{\percent}) of these  mechanisms are longer than $l=6$ steps. We therefore decided to limit our search to mechanisms of length $l\leq 6$, thus limiting the state space expansion to a depth of $3$.

For $786$ (\SI{8.9}{\percent}) of the $8805$ Rhea reactions we find at least one mechanism with $l\leq 6$ that is catalyzed exclusively by a single amino acid (henceforth sAA mechanism). For another $156$ (\SI{1.8}{\percent}) reactions, we find a completely non-catalytic mechanism that relies on no active amino acids. While the chemical validity of such proposals cannot be immediately dismissed, the fact that we find non-catalytic mechanisms suggests that some of our rules might lack context that would restrict their application to substrates that have been previously activated by interaction with a catalyst. Finally, for $169$ reactions we were unable to determine whether an sAA mechanism with $l\leq 6$ can be proposed, because the state space could not be explored within the allocated time ($180$ seconds on a standard laptop computer).

Only $8$ out of the $17$ amino acids considered were used by at least one of the constructed sAA mechanisms. Two of these amino acids, arginine and tyrosine, are used in $5$ and one mechanism, respectively. For these mechanisms we also found alternatives that operate with a different amino acid (in the case of arginine: aspartate and histidine; and in the case of tyrosine: histidine and lysine). Among the remaining $6$ amino acids we observed predominantly common proton acceptors and donors, such as aspartate, glutamate and histidine, while cysteine, lysine and threonine were used only rarely. The number of reactions solvable by one of the $6$ amino acids is given in \hyperref[tab:used_amino_acid]{Table~\ref{tab:used_amino_acid}}.

Glutamate is the most commonly used amino acid in our mechanisms, but our procedure also generated alternatives for the same reaction that rely on some other amino acid. For example, alternative mechanisms using histidine were found for $387$ reactions and $102$ reactions were catalyzed by aspartate. Interestingly, only $5$ reactions were found to be catalyzed by both aspartate and histidine; these were also reproduced by arginine-based mechanisms.

The disparity between the extents to which individual amino acids are used can be explained in part by the frequency with which these amino acids appear in the M-CSA. Aspartate, glutamate and histidine are indeed the three most common amino acids among all M-CSA mechanisms as well as among the covered sAA mechanisms. Moreover, the $8$ amino acids most commonly deployed by M-CSA mechanisms correspond almost to the $8$ amino acids utilized by the Rhea mechanisms we generated with our procedure. Only serine, the $7$\textsuperscript{th} most common amino acid in the M-CSA, does not show up in any of our mechanisms; its role is taken by threonine. However, the disparity in usage among amino acids that occur in our mechanisms is much more pronounced than in the M-CSA. This suggests that our rule set favors mechanisms that limit the interaction with the amino acid to proton transfers.

Despite the limitation on length and the restriction to catalysis by a single amino acid, our results include some interesting mechanisms. While many of the generated mechanisms consist of rules that were all derived from the same M-CSA mechanism, others contain rules derived from multiple distinct M-CSA mechanisms. For instance, a two-step glutamate-based mechanism that combines rules abstracted from distinct M-CSA mechanisms could be constructed for $54$ reactions, such as the example in \hyperref[fig:Crotonobetainyl-CoA_hydratase]{Figure~\ref{fig:Crotonobetainyl-CoA_hydratase}}. While glutamate itself acts only as a proton acceptor, the rules are utilized in nontrivial chemistry, consisting of an assisted keto-enol tautomerisation followed by unimolecular elimination and dehydration.

The mechanism in \hyperref[fig:Crotonobetainyl-CoA_hydratase]{Figure~\ref{fig:Crotonobetainyl-CoA_hydratase}} is but one example of amino acids engaging only in proton transfers yet triggering a more complex interaction. For instance, in $517$ Rhea reactions with a rule-based mechanism (\SI{67.3}{\percent}) the mechanism involves the creation or dissolution of a carbonyl group, which is common in organic chemistry. One of them is shown in \hyperref[fig:Crotonobetainyl-CoA_hydratase]{Figure~\ref{fig:Crotonobetainyl-CoA_hydratase}}.

\subsection{A More Complex Example}
\label{sec:example}

\begin{figure*}
	\centering
	\includegraphics[width=\textwidth]{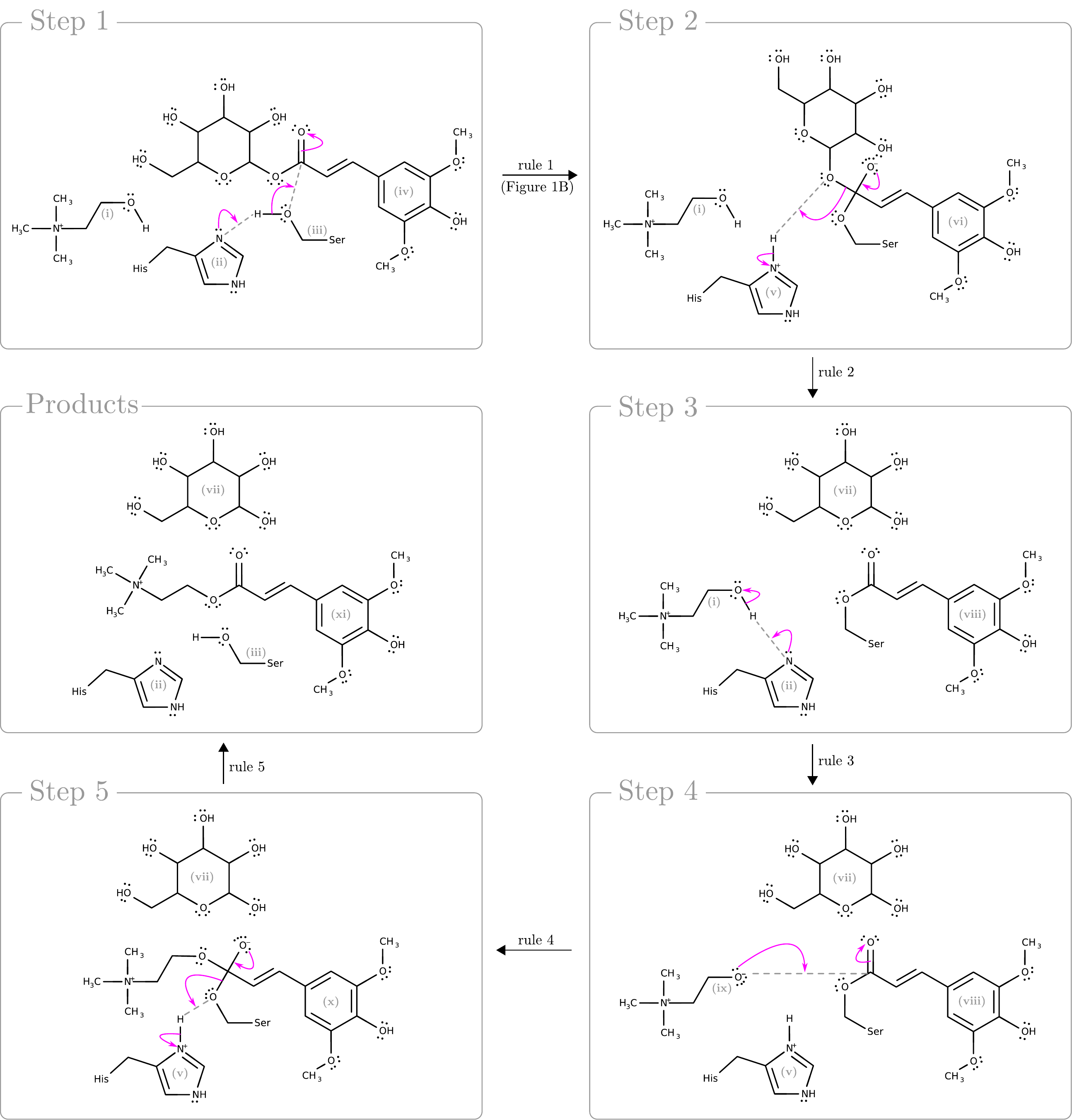}
	\caption{%
	Proposed 5-step mechanism for the conversion of choline \textit{(i)} and sinapoyl-glucose \textit{(iv)} into glucose \textit{(vii)} and sinapoyl-choline \textit{(xi)} (\rhea{12024} entry) depending on two amino acids, namely histidine \textit{(ii)} and serine \textit{(iii)}. %
	In the initial step histidine (ii) acts as proton acceptor of the hydrogen atom released from serine (iii). %
	Said serine (iii) then acts as a nucleophile towards the ester of the sinapoyl-glucose (iv) which results in the covalent linkage of the former to the latter. %
	In the second step the formed oxyanion collapses which causes the deprotonation of a fully protonated histidine (v) and elimination of a glucose molecule (vii). %
	The third step describes the deprotonation of choline (i) by histidine (ii). %
	In step four the activated choline (ix) attacks the transition molecule (viii) created during step 2. %
	In step 5, after covalent binding of choline (ix), the oxyanion in the transition molecule (x) collapses which results in the deprotonation of histidine (v), the release of serine (iii), and ultimately the formation of  sinapoyl-choline (xi). %
	Electronic displacements are shown as arrow pushes in magenta. %
	Arrows between the Step panels indicate which rule was applied. %
	Rule 1 is depicted in \hyperref[fig:rule_derivation]{Figure \ref{fig:rule_derivation}B}. %
	The states of the molecules are indicated by the roman numerals (i) through (x). %
	}
	\label{fig:2aminos}
\end{figure*}

Conducting a comprehensive exploration to seek mechanisms using more than one amino acid must be left to future work. Just trying any combination of two amino acids requires $26^2 = 676$ state space constructions and trace searches for each of the $8805$ eligible reactions in the Rhea database. Instead, we demonstrate that we can query for mechanisms that exhibit a specified behavioral motif found in the M-CSA by carefully selecting the set of catalytic amino acids when expanding the state space.

One such behavior motif that is common among several of the covered M-CSA mechanisms consists of the joint action of histidine and serine (or cysteine). Specifically, histidine acts as a proton acceptor depronotating serine (or cysteine) thereby activating the latter and allowing it to attack the substrate in a nucleophilic addition, which results in the formation of a covalent enzyme-substrate complex. We can ask whether the same behavior can be used in the construction of catalytic mechanisms (based on our rule set) for Rhea reactions. 

We thus define the set of catalytic amino acids to consist of histidine, cysteine, and serine, including all their tautomers as indicated in the previous section. We then search for mechanisms limited in length to less or equal $6$ steps within the state space $S_6$ for each Rhea reaction for which our procedure could not find any sAA mechanism.

For $133$ of the $7863$ ($= 8805 - 786 - 156$) target reactions, we find a mechanism conforming to the above criteria; for $69$ reactions our procedure timed out and for the remaining ones no such mechanism was found.

As an example of a reaction that can be catalyzed by the sought-after mechanism, consider the Rhea reaction \rhea{12024}. For this reaction, we find $173$ mechanisms each using $2$ to $5$ active amino acids. Only $2$ of these use exactly one histidine and one serine. One of the identified mechanisms is depicted in \hyperref[fig:2aminos]{Figure~\ref{fig:2aminos}}. The reaction consists of converting \iupac{1-\oxygen-(trans-sinapoyl)-\b-\dextro-glucose} and choline into \iupac{\dextro-glucose} and \iupac{\oxygen-sinapoyl-choline}. As the associated EC number of the reaction (\ECnum{2}{3}{1}{91}) already suggests, the \iupac{\dextro-glucose} moiety of \iupac{1-\oxygen-(trans-sinapoyl)-\b-\dextro-glucose} is replaced by a choline molecule.

The proposed reaction mechanism can be split into two parts, each consisting of an addition and a subsequent elimination step. The first part of the mechanism is based on two rules, of which the first is depicted in \hyperref[fig:rule_derivation]{Figure~\ref{fig:rule_derivation}B}. In step 1 the enzyme covalently binds to the substrate via a single bond between serine and sinapoyl-glucose, resulting in the formation of an oxyanion on the substrate side. The nucleophilic attack of the substrate by the serine is facilitated by the action of histidine, which abstracts a proton from the serine. The second rule application (step 2) yields the collapse of the unstable oxyanion. Together with the fully protonated histidine, the collapse results in the release of the glucose molecule from the enzyme-substrate complex. At the current stage of development our model does not allow differentiation between stereo-isomers and we therefore cannot identify specifically \dextro-glucose as listed in \rhea{12024}

The second part of the proposed mechanism is based on three different rules, but follows a very similar pattern as the first part just described. In step 3, the choline is deprotonated by a histidine, which allows the deprotonated choline to attack the enzyme-substrate complex in the fourth step 4, resulting in the covalent attachment of the choline to the enzyme-substrate complex. In the final and fifth step 5, the oxyanion formed in the process collapses, and with assistance from a proton transfer by the fully protonated histidine, eliminates the serine from the enzyme-substrate complex. This final step results in the catalytic components being restored and the products formed. 

The two rules used in the first part of the hypothetical mechanism have been extracted from four different M-CSA mechanisms in which they jointly occur (entries number $337$, $705$, $733$ and $866$). In contrast, the three rules comprising the second part are each derived from a different M-CSA mechanism. Thus, the constructed mechanism combines knowledge from at least four different mechanisms listed in the M-CSA.

The constructed mechanism employs a histidine for two distinct tasks. Histidine acts as a proton acceptor twice; it is restored in the middle of the mechanism after the first use and reused a second time thereafter. This suggests the possibility that a different proton acceptor, e.g.\@ a second instance of a histidine, could be deployed in one of the addition / elimination steps, should the catalytic site geometry of an actual enzyme require it.

The other mechanism that uses one serine and one histidine is functionally equivalent to the described mechanism, except that the histidine-enabled attachment of the serine to the substrate occurs in two steps; specifically, via step 1 and 2
of \mcsaid{94} proposal 1.

Among the $173$ mechanisms constructed for \rhea{12024} we identified mechanisms with the same chemistry as presented in \hyperref[fig:2aminos]{Figure~\ref{fig:2aminos}} but using cysteine in place of serine to anchor the substrate. We chose to detail the serine example since aspartate, histidine and serine are predicted by similarity to be (part of) the active site of the only protein, \uniprot{Q8VZU3}, listed as playing a role in \rhea{12024}. The presence of aspartate, in addition to histidine and serine, in the active site is interesting as in all of the relevant M-CSA mechanisms, histidine is assisted by aspartate or, in one case, glutamate as a passive amino acid. This catalytic triad engages in a fairly common process in enzymatic reactions. A hydrogen bond between aspartate (or glutamate) and histidine is increasing the \pKa~ of the latter \citep{Stehle2006}, thus expediting the deprotonation of serine and choline. As aspartate (or glutamate) is not part of the reaction center, it is not present in the rules derived from these cases.


\section{Discussion}
\label{sec:discussion}

Enzymatic catalysis is critical for enabling efficient and cost-effective chemical synthesis for medical, environmental, and industrial applications. The design of enzymes is therefore highly desirable. Among the numerous aspects that must be taken into account when designing an enzyme is the draft of a catalytic mechanism: a sequence of steps that is cyclical in the participating amino acids and whose traversal converts substrate(s) into product(s). Each step leads to a transition intermediate and contributes to the requirements that the architecture of the catalytic site must satisfy to make the mechanism effective. Such requirements include spatial arrangement and the fashioning of a physicochemical milieu.

In this paper we demonstrate that the drafting of mechanisms can be approached computationally. Central to our approach is the well-defined formalism of graph transformation with which we encode chemistry in the form of rules used to generate reactions by directly rewriting chemical graphs representing molecules. The approach thus hinges on extracting rules of chemistry from reaction examples, effectively generalizing these examples. Once rules are available they can be deployed to construct chemical spaces in which to search for suitable mechanisms.

We show the feasibility of this approach by converting the M-CSA into rules pertinent to amino acid side chain chemistry. In this way we make the knowledge in the M-CSA executable in the sense of enabling the computational construction of hypothetical mechanisms that catalyze reactions outside its scope. 

Specifically, we construct multiple mechanisms using a single amino acid to catalyze a large number of reactions in the Rhea database. The analysis of these mechanisms indicates that we succeed in capturing interesting and meaningful chemistry resulting from the combination of rules derived from reaction steps belonging to distinct M-CSA mechanisms.

Our procedure also generates a hypothetical catalytic mechanism relying on two amino acids---serine and histidine---for the Rhea reaction of sinapoyl-glucose and choline into glucose and sinapoyl-choline. For this reaction the literature \citep{Fraser2007} suggests an enzyme (\uniprot{Q8VZU3}) whose active site is predicted to include Ser178, Asp389 and His443 based on similarity. Aspartate is known to influence the \pKa~of histidine \citep{Stehle2006} when in proximity, thus playing a \enquote{passive} role (\hyperref[sec:catalysis]{Section \ref{sec:catalysis}}), which prevents it from being included in our rules as they are presently constructed (\hyperref[sec:rule_contruction]{Section \ref{sec:rule_contruction}}). The example leads us to believe that in addition to proposing novel catalytic mechanisms for general reactions, our approach is capable to assist in the prediction of catalytic sites of enzymes with known substrates and products but unknown or incomplete mechanisms. 

The generality of rules can be regimented by tuning the context of their action (\hyperref[sec:rule_contruction]{Section \ref{sec:rule_contruction}}).  Depending on the extent of context, a varying number of mechanisms can be subsumed under the same set of rules from which they can be generated, much like a formal language. It is conceivable therefore to use explicit rule compositions to formulate search criteria for retrieving mechanisms of specified chemistry from mechanism databases.

There are many possibilities for advancing the expressivity of rules as implemented by MØD, most notably by inclusion of stereochemistry \citep{Andersen2017} and the decoration of rules with application constraints taking physicochemical parameters into consideration. The navigation of large state spaces is a general challenge in computational science, but the results we obtained with the simple rule-heuristics employed here suggest that the automated generation of catalytic mechanisms for arbitrary reactions is a meaningful goal to pursue.

\section*{Acknowledgments}
The authors gratefully acknowledge Leon Middelboe Hansen and Mikkel Pilegaard for providing scripts constituting the preliminary analysis of the rules. Bernhard Thiel for his insightful pre-study during his master project at the University of Vienna.

\section*{Funding}
This work is supported by the Novo Nordisk Foundation grant NNF19OC0057834 and by the Independent Research Fund Denmark, Natural Sciences, grants DFF-0135-00420B and DFF-7014-00041.

\bibliographystyle{abbrvnat}
\bibliography{manuscript}

\begin{thebibliography}{21}
\providecommand{\natexlab}[1]{#1}
\providecommand{\url}[1]{\texttt{#1}}
\expandafter\ifx\csname urlstyle\endcsname\relax
  \providecommand{\doi}[1]{doi: #1}\else
  \providecommand{\doi}{doi: \begingroup \urlstyle{rm}\Url}\fi

\bibitem[Andersen et~al.(2013)Andersen, Flamm, Merkle, and
  Stadler]{AndersenFMS2013}
J.~L. Andersen, C.~Flamm, D.~Merkle, and P.~F. Stadler.
\newblock Inferring chemical reaction patterns using rule composition in graph
  grammars.
\newblock \emph{Journal of Systems Chemistry}, 4\penalty0 (1):\penalty0 4,
  2013.
\newblock ISSN 1759-2208.
\newblock \doi{10.1186/1759-2208-4-4}.

\bibitem[Andersen et~al.(2014)Andersen, Flamm, Merkle, and Stadler]{strat:14}
J.~L. Andersen, C.~Flamm, D.~Merkle, and P.~F. Stadler.
\newblock Generic strategies for chemical space exploration.
\newblock \emph{International Journal of Computational Biology and Drug
  Design}, 7\penalty0 (2/3):\penalty0 225--258, 2014.
\newblock \doi{10.1504/IJCBDD.2014.061649}.

\bibitem[Andersen et~al.(2016)Andersen, Flamm, Merkle, and Stadler]{mod}
J.~L. Andersen, C.~Flamm, D.~Merkle, and P.~F. Stadler.
\newblock {A software package for chemically inspired graph transformation}.
\newblock In R.~Echahed and M.~Minas, editors, \emph{Graph Transformation. ICGT
  2016. Lecture Notes in Computer science}, volume 9761, pages 73--88.
  Springer, Cham, Switzerland, 2016.
\newblock ISBN 978-3-319-40529-2.
\newblock \doi{10.1007/978-3-319-40530-8_5}.

\bibitem[Andersen et~al.(2017)Andersen, Flamm, Merkle, and
  Stadler]{Andersen2017}
J.~L. Andersen, C.~Flamm, D.~Merkle, and P.~F. Stadler.
\newblock {Chemical graph transformation with stereo-information}.
\newblock In J.~de~Lara and D.~Plump, editors, \emph{Graph Transformation. ICGT
  2017. Lecture Notes in Computer Science}, volume 10373, pages 54--69.
  Springer, Cham, Switzerland, 2017.
\newblock ISBN 978-3-319-61469-4.
\newblock \doi{10.1007/978-3-319-61470-0_4}.

\bibitem[Cook et~al.(2012)Cook, Johnson, Law, Mirzazadeh, Ravitz, and
  Simon]{Cook:2012}
A.~Cook, A.~P. Johnson, J.~Law, M.~Mirzazadeh, O.~Ravitz, and A.~Simon.
\newblock Computer-aided synthesis design: 40 years on.
\newblock \emph{WIREs Computational Molecular Science}, 2:\penalty0 79--107,
  2012.
\newblock \doi{10.1002/wcms.61}.

\bibitem[Ehrig(1979)]{10.1007/BFb0025714}
H.~Ehrig.
\newblock Introduction to the algebraic theory of graph grammars (a survey).
\newblock In V.~Claus, H.~Ehrig, and G.~Rozenberg, editors,
  \emph{Graph-Grammars and Their Application to Computer Science and Biology},
  pages 1--69, Berlin and Heidelberg, Germany, 1979. Springer.
\newblock ISBN 978-3-540-09525-5.
\newblock \doi{10.1007/BFb0025714}.

\bibitem[Ehrig et~al.(1973)Ehrig, Pfender, and Schneider]{EhrigPS1973}
H.~Ehrig, M.~Pfender, and H.~J. Schneider.
\newblock Graph-grammars: An algebraic approach.
\newblock In \emph{14th Annual Symposium on Switching and Automata Theory
  ({SWAT} 1973)}, pages 167--180, USA, 1973.
\newblock \doi{10.1109/SWAT.1973.11}.

\bibitem[Ehrig et~al.(2006)Ehrig, Ehrig, Golas, and Taentzer]{EhrigEGT2006}
H.~Ehrig, K.~Ehrig, U.~Golas, and G.~Taentzer.
\newblock Fundamentals of algebraic graph transformation.
\newblock In W.~Brauer, G.~Rozenberg, and A.~Salomaa, editors, \emph{Monographs
  in Theoretical Computer Science. An EATCS Series}. Springer, Berlin and
  Heidelberg, Germany, 2006.
\newblock ISBN 3-540-31187-4.
\newblock \doi{10.1007/3-540-31188-2}.

\bibitem[Fraser et~al.(2007)Fraser, Thompson, Shirley, Ralph, Schoenherr,
  Sinlapadech, Hall, and Chapple]{Fraser2007}
C.~M. Fraser, M.~G. Thompson, A.~M. Shirley, J.~Ralph, J.~A. Schoenherr,
  T.~Sinlapadech, M.~C. Hall, and C.~Chapple.
\newblock {Related Arabidopsis serine carboxypeptidase-like sinapoylglucose
  acyltransferases display distinct but overlapping substrate specificities}.
\newblock \emph{Plant Physiology}, 144\penalty0 (4):\penalty0 1986--1999, 8
  2007.
\newblock ISSN 0032-0889.
\newblock \doi{10.1104/pp.107.098970}.

\bibitem[Habel et~al.(2001)Habel, M{\"u}ller, and Plump]{HabelMP2001}
A.~Habel, J.~M{\"u}ller, and D.~Plump.
\newblock Double-pushout graph transformation revisited.
\newblock \emph{Mathematical Structures in Computer Science}, 11\penalty0
  (5):\penalty0 637--688, 2001.
\newblock \doi{10.1017/S0960129501003425}.

\bibitem[Hagberg et~al.(2008)Hagberg, Schult, and Swart]{networkx}
A.~A. Hagberg, D.~A. Schult, and P.~J. Swart.
\newblock Exploring network structure, dynamics, and function using networkx.
\newblock In G.~Varoquaux, T.~Vaught, and J.~Millman, editors,
  \emph{Proceedings of the 7th Python in Science Conference}, pages 11--15,
  Pasadena, CA USA, 2008.

\bibitem[Lombardot et~al.(2019)Lombardot, Morgat, Axelsen, Aimo,
  Hyka-Nouspikel, Niknejad, Ignatchenko, Xenarios, Coudert, Redaschi, and
  Bridge]{Lombardot2019}
T.~Lombardot, A.~Morgat, K.~B. Axelsen, L.~Aimo, N.~Hyka-Nouspikel,
  A.~Niknejad, A.~Ignatchenko, I.~Xenarios, E.~Coudert, N.~Redaschi, and
  A.~Bridge.
\newblock {Updates in Rhea: SPARQLing biochemical reaction data}.
\newblock \emph{Nucleic Acids Research}, 47\penalty0 (D1):\penalty0 D596--D600,
  1 2019.
\newblock ISSN 0305-1048.
\newblock \doi{10.1093/nar/gky876}.

\bibitem[Pleissner and K{\"u}mmerer(2020)]{Pleissner:2020}
D.~Pleissner and K.~K{\"u}mmerer.
\newblock Green chemistry and its contribution to industrial biotechnology.
\newblock In M.~Fröhling and M.~Hiete, editors, \emph{Advances in Biochemical
  Engineering/Biotechnology}, volume 173, pages 281--298, Cham, Switzerland,
  2020. Springer.
\newblock ISBN 978-3-030-47065-4.
\newblock \doi{10.1007/10\_2018\_73}.

\bibitem[Ribeiro et~al.(2017)Ribeiro, Holliday, Furnham, Tyzack, Ferris, and
  Thornton]{RibeiroHFTFT2017}
A.~J.~M. Ribeiro, G.~L. Holliday, N.~Furnham, J.~D. Tyzack, K.~Ferris, and
  J.~M. Thornton.
\newblock {Mechanism and Catalytic Site Atlas (M-CSA): a database of enzyme
  reaction mechanisms and active sites}.
\newblock \emph{Nucleic Acids Research}, 46\penalty0 (D1):\penalty0 D618--D623,
  2017.
\newblock ISSN 0305-1048.
\newblock \doi{10.1093/nar/gkx1012}.

\bibitem[Schrittwieser et~al.(2018)Schrittwieser, Velikogne, Hall, and
  Kroutil]{Schrittwieser:2018}
J.~H. Schrittwieser, S.~Velikogne, M.~Hall, and W.~Kroutil.
\newblock Artificial biocatalytic linear cascades for preparation of organic
  molecules.
\newblock \emph{Chemical Reviews}, 118\penalty0 (1):\penalty0 270--348, 2018.
\newblock \doi{10.1021/acs.chemrev.7b00033}.

\bibitem[Segler and Waller(2017)]{Segler:2017}
M.~H.~S. Segler and M.~P. Waller.
\newblock Modelling chemical reasoning to predict and invent reactions.
\newblock \emph{Chemistry - A European Journal}, 23\penalty0 (25):\penalty0
  6118, 2017.
\newblock \doi{10.1002/chem.201604556}.

\bibitem[Snider and Wolfenden(2000)]{Snider2000}
M.~J. Snider and R.~Wolfenden.
\newblock {The rate of spontaneous decarboxylation of amino acids}.
\newblock \emph{Journal of the American Chemical Society}, 122\penalty0
  (46):\penalty0 11507--11508, 11 2000.
\newblock ISSN 0002-7863.
\newblock \doi{10.1021/ja002851c}.

\bibitem[Stehle et~al.(2006)Stehle, Brandt, Milkowski, and Strack]{Stehle2006}
F.~Stehle, W.~Brandt, C.~Milkowski, and D.~Strack.
\newblock {Structure determinants and substrate recognition of serine
  carboxypeptidase-like acyltransferases from plant secondary metabolism}.
\newblock \emph{FEBS Letters}, 580\penalty0 (27):\penalty0 6366--6374, 11 2006.
\newblock ISSN 00145793.
\newblock \doi{10.1016/j.febslet.2006.10.046}.

\bibitem[Todd(2005)]{Todd:2005}
M.~H. Todd.
\newblock Computer-aided organic synthesis.
\newblock \emph{Chemical Society Reviews}, 34:\penalty0 247--266, 2005.
\newblock \doi{10.1039/b104620a}.

\bibitem[Welborn and Head-Gordon(2019)]{Welborn:2019}
V.~V. Welborn and T.~Head-Gordon.
\newblock Computational design of synthetic enzymes.
\newblock \emph{Chemical Reviews}, 119\penalty0 (11):\penalty0 6613--6630,
  2019.
\newblock \doi{10.1021/acs.chemrev.8b00399}.

\bibitem[Zimmerman et~al.(2020)Zimmerman, Anastas, Erythrope, and
  Leitner]{Zimmerman:2020}
J.~B. Zimmerman, P.~T. Anastas, H.~C. Erythrope, and W.~Leitner.
\newblock Designing for a green chemistry future.
\newblock \emph{Science}, 367\penalty0 (6476):\penalty0 397--400, 2020.
\newblock \doi{10.1126/science.aay3060}.

\end{thebibliography}

\end{document}